\documentclass[prb,twocolumn,amsfonts,amssymb,amsmath,aps]{revtex4}

\usepackage{graphicx}
\usepackage{bm}
\usepackage{psfig}

\begin{document}

\title{Inhomogeneously doped two-leg ladder systems}

\author{Stefan Wessel}
\author{Martin Indergand}
\author{Andreas L\"auchli}
\author{Urs Ledermann}
\author{Manfred Sigrist}

\affiliation{Theoretische Physik, ETH-H\"onggerberg, CH-8093 Z\"urich, Switzerland}

\date{\today}

\begin{abstract}
A chemical potential difference between the legs of a two-leg ladder is found
to be harmful for Cooper pairing. The instability of superconductivity in such
systems is analyzed by compairing results of various analytical and numerical
methods.
Within  a
strong coupling approach for the $t$-$J$ model, supplemented by exact numerical
diagonalization, 
hole binding is found  unstable 
beyond a finite, critical chemical potential difference.
The spinon-holon mean field theory for the $t$-$J$ model shows a clear
reduction of the the BCS gaps upon increasing the chemical potential
difference leading to a breakdown of superconductivity.
Based on a renormalization group approach and Abelian bosonization,
the doping dependent phase diagram 
for the weakly interacting Hubbard model with different chemical potentials  was
determined. 

\end{abstract}

%\pacs{}

\maketitle 
\section{Introduction}\label{introduction}
Doped spin liquids have
been an important subject of condensed matter research for the last two
decades, mainly due to their possible relevance to the (cuprate)
high-temperature superconductors (HTSCs).\cite{Anderson1,FiRev}
Although it is still not understood how high-temperature superconductivity
emerges from an antiferromagnetic Mott-insulator upon carrier doping,
there is broad consensus that an intermediate pseudogap phase plays
a crucial role in understanding both the exotic normal and
superconducting properties of these materials.\cite{TiSt}

Many proposals concerning the nature of the pseudogap phase have emerged.
One candidate for
the pseudogap phase is the resonating valence bond state
(RVB), which describes strongly fluctuating short-ranged spin singlets.\cite{Anderson1}
While the relevance of this state to the quasi-two-dimensional HTSCs is still
under debate, it is
realized in specially designed lattice structures.\cite{CTTT} Of
particular interest are systems with  ladder-shaped crystal
structures, which are realized in
various transition metal oxide compounds.\cite{AHT,scienceDR}

Undoped, these ladder-systems are Mott-insulators and well described by
quantum spin-1/2 Heisenberg models. For the two-leg ladder the ground
state is  an RVB-like state displaying short-ranged spin-singlet
correlations with spin-singlet dimers on the rungs dominating over those
along the legs. This constitutes a spin liquid  with a finite
energy gap to the lowest spin excitations. Furthermore, it was proposed that
such a system would exhibit
superconductivity upon hole-doping.\cite{DRS2L,RGS2L} Various theoretical approaches
confirmed a strong tendency towards formation of Cooper pairs with
phase properties reminding
of the $d_{x^2-y^2}$-wave channel of the two-dimensional
HTSCs.\cite{HPNSH,troyer,balents,LeHur}

The theoretical proposals were followed by an intense material research
attempting to dope holes into a variety of known insulating 
copper-oxide compounds
displaying ladder structures.\cite{DagottoRev} Materials under consideration
are SrCu$_2$O$_3$, CaCu$_2$O$_3$, LaCuO$_{2.5}$, and
Sr$_{14}$Cu$_{24}$O$_{41}$. Doping these compounds is intrinsically
difficult because of chemical instabilities, and 
carrier localization effects may inhibit the desired metallic behavior.
Nevertheless, the search for superconductivity has been successful in the
compound Sr$_{14-x}$Ca$_x$Cu$_{24}$O$_{41}$
which contains layers of ladders alternating with layers of single
chains.\cite{pressure} In this material $T_c$ rises up  to about 12 K under high pressure, which
apparently leads to a transfer of charge carriers from the chains onto the ladders.
However, a detailed understanding of this system under
high pressure has not been reached yet. A well-controlled and less
invasive way of doping a ladder compound is thus highly desirable.

Hole and electron doping  
by means of field effect devices
induces  mobile charge carriers 
into the originally insulating 
material using a large gate voltage. This method
would be ideal for doping quasi one- and two-dimensional 
systems, since the induced charge is confined to the outer-most layer of the
compound, closest to the gate. 

 An alternative
technique of tunable doping has been achieved using heterostructures of
layered materials such high-$T_c $ -cuprates combined with
ferroelectrics like Pb(Zr,Ti)O$_3$ \cite{Triscone1,Triscone2,Triscone3}.
These non-volatile techniques 
of tunable doping may allow for a detailed comparison between
experiment 
and theory in these low-dimensional structures.

The most natural choice of a ladder system for this type of doping
controll is a
film in which the ladder planes lie parallel to the gate or
ferroelectric, so that the 
carriers enter the ladders uniformly. However, in the compound LaCuO$_{2.5}$ the
ladders are not parallel to each 
other, but exhibit a staggering.\cite{HiTak} Consequently,  in a field effect
device the ladders would be inhomogeneously doped in
the sense that the chemical potential on the two legs would be
different. Similarly, a variable orientation of the dipolar moments of the
ferroelectric can lead to inhomogeneous doping. 

In this paper we analyze the evolution of the 
superconducting state under such doping circumstances. Our analytical
and numerical  analysis shows 
that non-uniform doping is harmful to  the superconducting state of
the two-leg ladder.
 While for small
$\Delta\mu$ the ladder remains superconducting, the pairing is
suppressed  upon increasing 
$\Delta\mu$, and depending on the
doping level new phases with reduced, and eventually without
superconductivity appear.  

The paper is organized according to the used analytical and numerical
methods. In this way we study various aspects of the problem within
different approximative schemes. 
In the following section a qualitative argument for the limited stability of
the superconducting state upon inhomogeneous doping is presented. 
In Sec.~III,
a discussion of numerical exact diagonalization results is given for
the charge correlations of two holes in ladders with up to 22 sites.
Then, in Sec.~IV we consider a mean field treatment of the $t$-$J$
model based on the spinon-holon decoupling scheme. In Sec.~V, we apply  
renormalization group (RG) and bosonization methods to derive the
phase diagram of the 
weakly interacting Hubbard model in the inhomogeneously doped case.
We conclude in Sec. VI and draw a unified picture of the behavior of
inhomogeneously doped two-leg ladder systems by combining the results from
the earlier sections.

\section{Strong rung coupling limit}\label{strongcoupling}
The influence of a difference in the
chemical potential on the pairing state of the two-leg ladder can be
illustrated by a simple qualitative argument for the $t$-$J$ model. 
Consider the two-leg ladder with electrons moving along
the legs and rungs with hopping matrix elements $t$ 
and $t'$ respectively, and  nearest-neighbor spin exchange with
exchange constants $J$ and $J'$. The Hamiltonian of the $t$-$J$ model
then reads 
\begin{eqnarray}\label{hamiltonian}
H&=&-t\sum_{j,a,s}{\cal P}(c^{\dag}_{j,a,s}c_{j+1,a,s}+\mathrm{h.c.}){\cal P}
\nonumber\\
&&-t'\sum_{j,s}{\cal P}(c^{\dag}_{j,1,s}c_{j,2,s}+\mathrm{h.c.}){\cal P}
\nonumber\\
&&+J\sum_{j,a}({\mathbf{S}}_{j,a}\cdot{\mathbf{S}}_{j+1,a}-
\frac{1}{4}n_{j,a}n_{j+1,a})
\nonumber \\
&&+J'\sum_j({\mathbf{S}}_{j,1}\cdot{\mathbf{S}}_{j,2}
-\frac{1}{4}n_{j,1}n_{j,2} )
\nonumber\\
&&-\sum_{j,a,s} \mu_a \: n_{j,a,s}.
\end{eqnarray}
The operator $c^{\dag}_{j,a,s}$ ($ c_{j,a,s} $) creates (annihilates) an
electron with spin $s$ on site $(j,a)$, where $j$ labels the rungs, and
$a=1,2$ the legs. The electron number operators are defined as 
$n_{j,a,s}=c^{\dag}_{j,a,s}c_{j,a,s}$, and $n_{j,a}=\sum_s n_{j,a,s}$. The spin
operators are 
\begin{equation}
  {\bf S}^{\alpha}_{j,a}=\frac{1}{2}\sum_{s,s'}
  c^{\dag}_{j,a,s}{\sigma}^{\alpha}_{ss'}c_{j,a,s'},
\end{equation}
where $\sigma^{\alpha}$, $\alpha=1,2,3$ are  Pauli matrices.
The constraint of excluded double occupancy is enforced by the projection operator
\begin{equation}\label{constraint}
  {\cal P}=\prod_{j,a}(1 - n_{j,a,\uparrow}n_{j,a,\downarrow}).
\end{equation}
In the last line of Eq. (\ref{hamiltonian}) 
 different chemical potentials on the two legs,
$\mu_{a}$, $a=1,2$ describe an inhomogeneous doping of the system. Throughout
 this paper we assume $\Delta\mu=\mu_1-\mu_2\ge 0$, and refer to the leg  with
 $a=1$  ($a=2$) as the upper (lower) leg.
 Furthermore, the overall doping
 concentration $\delta=1-n$ fixes the average chemical potential
 $\bar{\mu}=(\mu_1+\mu_2)/2$.

The phases of the $t$-$J$ model on the two-leg ladder for
$\Delta\mu=0$ are well characterized.\cite{twoleg} 
At half-filling ($\delta=0$) with one electron per site  the
ladder is a (Mott-) insulating spin liquid. Upon removal of electrons, i.e. doping 
of holes, mobile carriers appear, resulting in a Luther-Emery liquid with gapless charge
modes and gapped spin excitations. Furthermore, the gapless charge mode
exhibits dominant superconducting correlations with d-wave-like phase structure.

We now discuss the effect of inhomogeneous doping on this
superconducting state,
described by $ \Delta \mu >0 $. For many aspects of ladder systems the limit of
strong rung-coupling gives useful insights into their basic properties. 
Therefore we first consider the Hamiltonian (1) in the limit $J',t'\gg J,t$. Neglecting the coupling along the
legs entirely, the 
undoped system corresponds to a chain of decoupled rungs, and the groundstate
becomes a product state of dimer 
spin-singlets on the rungs. Furthermore, the lowest  spin excitation corresponds to exciting one
rung-singlet to a triplet, at an energy expense of $J'$. The
superconducting state, i.e. Cooper pairing, in the doped spin liquid
is inferred from the fact that two doped holes rather reside on
a single rung rather than to separate onto two rungs. This is the case if the
dominant energy scale is the spin exchange interaction, $J'$. Then the cost
of breaking two spin singlets is larger than the gain of kinetic
energy from separating the two holes. Namely, for two holes on a single
rung the energy is 
\begin{equation}
  E_{2h}=J'-2 \bar{\mu},
\end{equation}
while for a single hole
\begin{equation}
  E_{1h}=J'-\bar{\mu}-\frac{1}{2}\sqrt{4(t')^2+\Delta\mu^2}.
\end{equation}
 Pairing on a rung is favored, if
\begin{equation}
  2 E_{1h}-E_{2h}=J'-\sqrt{4(t')^2+\Delta\mu^2}>0,
\end{equation}
which in the uniformly doped case  ($\Delta\mu=0$) leads to the condition
$J'>2t'$ for pairing. Obviously, a finite value of $ \Delta \mu $ weakens
the pairing by reducing the above energy gain. 

This simple observation of depairing under non-uniform doping 
is confirmed by more sophisticated approaches, as those considered in the following
sections.

\section{Exact Diagonalization}\label{numeric}
To extend the 
 discussion of the two-hole
problem beyond the strong coupling limit  we performed
 exact diagonalizations 
of finite systems, using the Lanczos algorithm.\cite{lanczos} 
We considered the Hamiltonian~(\ref{hamiltonian}) at isotropic coupling
($t=t'$, $J=J'$), and studied the half-filled system doped with two holes, using 
periodic boundary conditions.
We studied systems of different length, $L$, containing 8 to 11 rungs, and
furthermore considered different
values of $J/t$. 

Consistent with the strong coupling argument of Sec. II,  
the hole bound state is found to be unstable beyond a critical value
of $\Delta\mu>\Delta\mu_c$. Furthermore for the range of parameters considered
here
($0.4 \leq J/t \leq 0.8$), this critical value is $\Delta\mu_c \approx J'$ .
 This indicates that the physics of the
system is quite well captured 
 by the strong rung-coupling limit, with $J'$ being the dominant energy scale
 for pairing.

\begin{figure}[h]
\includegraphics[angle=270,width=\linewidth]{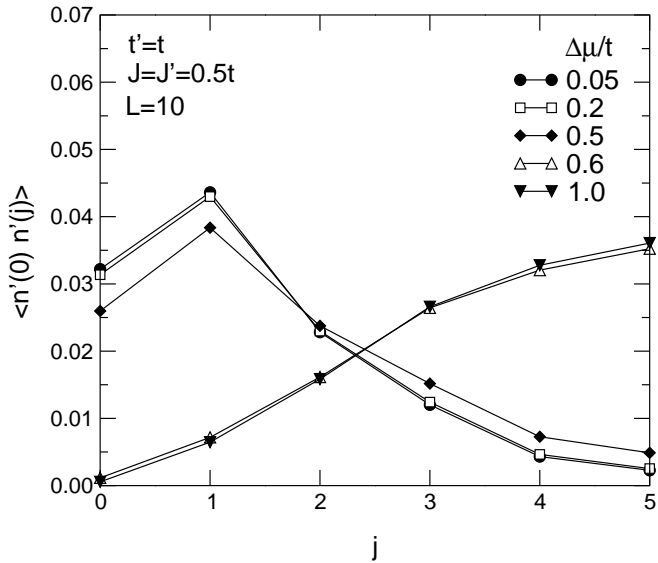}
\caption{\label{corrfunct} Rung hole-hole correlation-function,
  $\langle n'(0)n'(j) 
  \rangle$, in the ground state of the $t$-$J$ model for an  inhomogeneously doped two-leg ladder at
  selected values of the chemical potential difference, $\Delta\mu$. The finite
  ladder has $L=10$ rungs, and furthermore $t=t'$, and $J=J'=0.5t$.}
\end{figure}

The behavior of the holes under non-uniform doping can be analysed using the
hole-hole correlation function.
Denoting the hole number operator
on rung $j$ by $n'(j)=2-n_{j,1}-n_{j,2}$, the rung hole-hole correlation function
is defined as the ground state expectation value $\langle n'(0)n'(j) \rangle$,
for the rung-rung separation $j=0,1,\dots,\lfloor
\frac{L}{2} 
\rfloor$ . This
correlation function  is shown 
in Fig.~\ref{corrfunct} for a ladder with 10 rungs for $J/t=0.5$, and at selected
values of the chemical potential 
difference, $\Delta\mu$.
The behavior of the correlation function changes abruptly between $\Delta\mu/t=0.5$ and
$\Delta\mu/t=0.6$. For small values of $\Delta\mu/t\leq 0.5$, 
we find maximal rung hole-hole
correlations between neighboring rungs, and a strong decay of the
correlation function at larger distances. For values of
$\Delta\mu/t\geq 0.6$, the value of the 
correlation function is increasing with distance, and has a maximum value at
the maximal possible rung-rung separation. Furthermore, it almost vanishes on the same rung. This
clearly indicates the destruction of the hole-hole bound state for
$\Delta\mu/t\geq 0.6$, where
the system 
consists of two holes on the  lower
leg, favored by a lower chemical potential.
\\
The sudden change in the behavior of  the correlation function is due to a
level crossing at $\Delta \mu_c$ in this system. While in the bound regime the ground
state has zero total momentum along the legs, the lowest energy state  in the unbound regime
has a finite value of the total momentum.
The  correlation functions for $\Delta\mu/t= 0.2$ and $\Delta\mu/t= 0.05$ are  almost
identical, reflecting the robustness of the bound state against small doping inhomogenities. Indeed, it can be shown that
the doping asymmetry $\Delta\mu$  does not have
any effect in first order perturbation theory.
In the regime of unbound holes
the correlation function is again insensitive to changes in 
$\Delta\mu$, since for large $ \Delta \mu $ the holes are almost
exclusively located on the lower leg, and therefore 
increasing $\Delta\mu$  merely results in an overall energy shift.   

\begin{figure}
\includegraphics[angle=270,width=\linewidth]{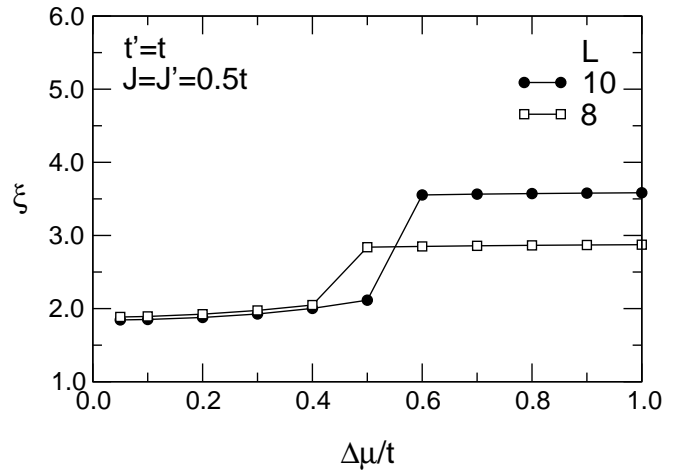}
\caption{\label{xi}Characteristic
  hole-hole separation length,
  $\xi$, 
in the ground state of the $t$-$J$ model for an  inhomogeneously doped two-leg ladder
as a function of the difference in
  the chemical potential, $\Delta\mu$, for $L=8$ (circles), and $L=10$ (squares).
  Furthermore, $t'=t$, and $J=J'=0.5t$.}
\end{figure}
The drastic change in the hole-hole correlation function as a result of the transition
between the bound and unbound regime is also reflected in 
 the characteristic hole-hole separation length, $\xi$, defined by 
\begin{equation}
  \label{eq:xi}
  \xi^2=\frac{1}{N}\sum_{j=-\lfloor\frac{L}{2}\rfloor}^{\lceil\frac{L}{2}\rceil-1}j^2 \: \langle  
  n'(0)n'(j) \rangle, 
\end{equation}
  where $N$ is a  normalization factor
 given by $N=\sum_{j} \langle n'(0)n'(j) \rangle$. In Fig.~\ref{xi}, this
 separation length is plotted as a function of
 $\Delta\mu$ for two 
 different systems  with an even number of rungs, $L=8$ and
 $L=10.$\cite{footnote}
For both systems the
 characteristic hole-hole separation length jumps discontinuously at a
 critical value of
 $\Delta\mu_c \approx J'$. Furthermore, in the bound regime, i.e\. for $\Delta\mu \leq
 \Delta\mu_c$,  finite size effects in this quantity
 are  rather small already  for the  system
 sizes considered here. In this regime the holes are bound, and the rung hole-hole correlation 
 decays exponentially. Therefore, the bound pair  wave function
 extends over only a few rungs.
  In the unbound regime however,
 the two holes tend to separate onto the lower leg, and therefore $\xi$ grows with increasing system size.

%The rung hole-hole correlation in Fig.~\ref{corrfunct} is the sum
%over the 4 site 
%hole-hole correlations $\langle n'_{i,a}n'_{i+j,b}\rangle \rangle$ with
%$a,b\in\{1,2\}$, where $n'_{i,a}$ is the hole number operator on site $(i,a)$
%given by $n'_{i,a}=1-n_{i,a}$. Fig.~\ref{enviro} shows all 
%hole-hole correlations between sites on the same rung and between sites on
%neighboring rungs as a function of $\Delta\mu$. The effect of increasing
%$\Delta\mu$ is a monotonic decrease of all hole-hole correlations except for
%$\langle n'_{i,2}n'_{i+1,2} \rangle$ which grows in the same way as   $\langle
%n'_{i,1}n'_{i+1,1} \rangle$ decreases. At $\Delta\mu_c$ all correlations would
%collapse to zero in the infinite system. The finite value of $\langle
%n'_{i,2}n'_{i+1,2} \rangle$ is due to the finite length, $L=10$, of our system.

%\begin{figure}[h]
%\includegraphics[angle=270,width=\linewidth]{./graphics/enviro.eps}
%\caption{\label{enviro}The hole-hole correlations $\langle
%  n'_{i,a}n'_{j,b}\rangle$ in the ground-state between sites on the same
% rung and between sites  on neighboring rungs as a function of the
% difference in the chemical potential $\Delta\mu/t$. $n'_{i,a}=1-n_{i,a}$, $L=10$,
% $J'=J=0.5t$, $t'=t$.}
%\end{figure}

The above numerical analysis demonstrates that non-uniform doping, by confining
the mobile carriers onto one of the legs, is harmful to pairing. Furthermore, the
interleg exchange interaction plays the important role of stabilizing the bound hole
pair state. 
While the finite ladders considered in this section may be viewed as systems
with a doping concentration of roughly $ \delta \sim  0.1 $,
different approaches  are needed  to analyze the  finite-doping
regime beyond the two-hole problem. These will be presented in the following sections.

\section{Mean field Analysis}\label{meanfield}
In this section we extend our analysis of the $t$-$J$-model by considering 
 a mean field approximation based on the 
spinon-holon-decoupling scheme. We follow a similar approach as  
various previous studies on ladders as well as two-dimensional
systems.\cite{threeleg,twoleg,plee,kotliar} 

\subsection{Spinon-Holon Decomposition}

The non holonomic local constraint $\sum_{s} c^{\dag}_{j,a,s} c_{j,a,s} \leq
1$ is one of the 
main difficulties in treating the $t$-$J$-model. The slave-boson formalism
provides a possibility to take this constraint into
account. Introducing  fermionic spinon operators $f$ and  bosonic
holon 
operators $b$, the electron creation and annihilation operators can be
expressed as   

\begin{equation} 
c^{\dag}_{j,a,s}=f^{\dag}_{j,a,s}b_{j,a}, \quad \textrm{and} \quad
c_{j,a,s}=b^{\dag}_{j,a}f_{j,a,s}, 
\end{equation}
leading to the  holonomic constraint 
\begin{equation}\label{holonomic}
\sum_{s} f^{\dag}_{j,a,s}f_{j,a,s} +b^{\dag}_{j,a}b_{j,a} = 1.
\end{equation}
The Hamiltonian  (\ref{hamiltonian}) can be expressed in terms of this new
operators as 
\begin{eqnarray}
H&=& -t  \sum_{j,a,s}\;
(f^{\dag}_{j,a,s}\,f_{j+1,a,s}\,b_{j,a}\,b^{\dag}_{j+1,a} +
\mathrm{h.c.})        \nonumber\\ 
&&-t'\sum_{j,s}\;
(f^{\dag}_{j,1,s}\,f_{j,2,s}\,b_{j,1}\,b^{\dag}_{j,2}   +
\mathrm{h.c.})  \nonumber\\ 
&&+J  \sum_{j,a}\;
%(b_{j,a} b^{\dag}_{j,a} b_{j+1,a} b^{\dag}_{j+1,a})\; 
{\mathbf{S}}^f_{j,a} \cdot {\mathbf{S}}^f_{j+1,a}
\nonumber\\ 
&&  +J'  \sum_{j} \;
%(b_{j,1} b^{\dag}_{j,1} b_{j,2} b^{\dag}_{j,2}) \;
{\mathbf{S}}^f_{j,1} \cdot
{\mathbf{S}}^f_{j,2}\nonumber\\ 
&& -  \sum_{j,a} \;  \lambda_{ja}\,\Big(\sum_{s}
f^{\dag}_{j,a,s}f_{j,a,s} + b^{\dag}_{j,a}b_{j,a} - 1\Big) \nonumber\\ 
&& +    \sum_{j,a} \; \mu_a \, b^{\dag}_{j,a}\,b_{j,a},
\label{shhamiltonian} 
\end{eqnarray}
where the Lagrange multipliers $\lambda_{ja}$, $a=1,2$ have been introduced  to
enforce the local 
constraint (\ref{holonomic}).
In the interaction part, the density-density terms $n_{j,a}\,
n_{j',a'}$ are omitted. Within the following mean field treatment,
this term would destroy the local $SU(2)$ gauge symmetry of the spinon
representation at half-filling.\cite{Affleck} This symmetry
corresponds to a local unitary rotation of the spinor $ (f_{i,a,
  \uparrow},  f^{\dag}_{i,a, 
  \downarrow}) $ leaving the spinon spectrum invariant.\cite{Affleck} This
symmetry appears naturally in the large-$U$ Hubbard model
\cite{Affleck} and is 
considered to be  essential for various aspects of the weakly doped
$t$-$J$-model.\cite{plee,honer} Therefore, we will keep only the spin
exchange part of the interaction which conserves this symmetry in
the mean field approximation.\cite{plee}
The 
last term in  Eq. (\ref{shhamiltonian}) takes  the different chemical
potentials on the two legs into account. 

To proceed we decouple the terms which are not single particle terms 
in the Hamiltonian
(\ref{shhamiltonian}) by introducing the following mean fields \cite{plee} 

\begin{eqnarray}\label{defmeanfields}
\chi_{j,a;j',a'} &=& \frac{1}{2} \sum_{s} \langle
f^{\dag}_{j,a,s}\,f_{j',a',s} \rangle, 
\nonumber\\
B_{j,a;j',a'} &=& \langle b_{j,a}\,b^{\dag}_{j',a'} \rangle,
\label{chibdel}\\[3mm] 
\Delta_{j,a;j',a'} &=& \langle f_{j,a,\downarrow}\,f_{j',a',\uparrow}
\rangle. \nonumber 
\end{eqnarray}
In the following the mean fields along the legs are labeled with
the indices 1 and 
2, and the mean field on the rung with the index 3, e.g. for $\chi$:
\begin{eqnarray}\label{meanfieldlabeling}
\chi_{a} &=& \frac{1}{2} \sum_{s} \langle
f^{\dag}_{j,a,s}\,f_{j+1,a,s} \rangle \qquad a=1,2 
\label{chileg}\\
\chi_{3} &=& \frac{1}{2} \sum_{s} \langle
f^{\dag}_{j,1,s}\,f_{j,2,s} \rangle. \nonumber 
\end{eqnarray}
This convention of labeling the bond $(j,a;j',a')$ also applies to the mean
fields $B$, and $\Delta$. Finally the doping level is fixed by the
condition 

\begin{equation}
1- \delta = \frac{1}{2}\sum_{a, s} \langle f^{\dag}_{j,a,s}
f_{j,a,s} \rangle \; .
\end{equation} 
The Lagrange multipliers $ \lambda_{ja} $ are kept uniform on each
leg, i.e. $ \lambda_{ja} \to \lambda_a $, so that the constraint is
satisfied only on the average on each leg of the ladder. 
%The four-boson operators in the exchange
%terms are replaced by their corresponding mean values which for leg bonds are
%\begin{equation}
%J \langle b_{j,a} b^{\dag}_{j,a} b_{j+1,a} b^{\dag}_{j+1,a} \rangle =
%J (1
%- B_a)^2 = J_a, \qquad 
%a =1,2 
%\end{equation}
%and along the rungs
%\begin{equation}
%J' \langle b_{j,1} b^{\dag}_{j,1} b_{j,2} b^{\dag}_{j,2} \rangle 
%= J'( 1 -2 \delta + B_3^2) = J_3 \; .
%\end{equation}
%These factors take the effective weakening of the exchange interaction due to the
% holes into account. Namely, in the presence of a hole the
%exchange interactions between the adjacent bonds become inactive.  

Introducing Fourier transformed operators
\begin{eqnarray}
f_{k,a,s} &=& \frac{1}{\sqrt{L}} \sum_j f_{j,a,s} \: e^{ikr_j},
 \nonumber\\
b_{k,a} &=& \frac{1}{\sqrt{L}} \sum_j b_{j,a}\:  e^{ikr_j},
\end{eqnarray}
the mean field Hamiltonian reads
\begin{widetext}
\begin{equation}\label{meanfieldham}
H_{MF} = \sum_k \Big(H^{b}_k + H^{f}_k\Big) +L\bigg[\sum_{a}
\Big\{\frac{3}{2} J\left(\Delta_{a}^{2} +\chi_{a}^{2} \right)  +4t\chi_aB_a \Big\} + \frac{3}{2} J_3 \left( \Delta_{3}^{2} +\chi_{3}^{2}
\right) +4t'\chi_3B_3 \bigg].
\end{equation}

The quadratic terms $H^{b}_k$ and $H^{f}_k$ are given by

\[
H^{b}_k = \left( \begin{array}{c} b^{\dag}_{k,1}\Big. \\ \Big.  b^{\dag}_{k,2}
\end{array}  \right)^{\top} \left[  \begin{array}{cc} -4t\chi_1\cos{k}
-\lambda_1 +\mu_1 &
-2t'\chi_3 \Big.\\\Big.   -2t'\chi_3 & -4t\chi_2\cos{k} -\lambda_2 +\mu_2 \end{array}  \right] \left( \begin{array}{c} b_{k,1} \Big.\\\Big. b_{k,2} \end{array}\!\! \right),
\]
\[
H^{f}_k = \left(\!\begin{array}{l} f^{\dag}_{k,1,\uparrow}\\
f^{\dag}_{k,2,\uparrow}\\ f_{-k,1,\downarrow}\\ f_{-k,2,\downarrow}
\end{array}\!\!\right)^{\!\!\top} \left[\begin{array}{cc} \; \hat{\xi}_k\; &
\hat{\Delta}_k \bigg. \\ \bigg. \hat{\Delta}_k & -\hat{\xi}_k \end{array} \right]\left( \!\begin{array}{l} f_{k,1,\uparrow} \\ f_{k,2,\uparrow} \\ f^{\dag}_{-k,1,\downarrow} \\ f^{\dag}_{-k,2,\downarrow} \end{array}\!\!\right),
\]     
where $\hat{\xi}_k$, and $\hat{\Delta}_k$ are the following $2\times2$ matrices
\[
\hat{\xi}_k = \left[ \begin{array}{cc} -\lambda_1 -\Big( 2tB_1 +
\frac{3}{2}J\chi_1\Big)\cos{k} &  -t'B_3 - \frac{3}{4}J_3\chi_3 \Big.\\\Big.
-t'B_3 - \frac{3}{4}J_3\chi_3 & -\lambda_2 -\Big( 2tB_2 +
\frac{3}{2}J\chi_2\Big)\cos{k} \end{array} \right], 
\qquad
\hat{\Delta}_k = \left[ \begin{array}{cc} \Big. -\frac{3}{2}J\Delta_1\cos{k}
& -\frac{3}{4}J\Delta_3 \Big.\\\Big.  -\frac{3}{4}J\Delta_3 &
-\frac{3}{2}J\Delta_2\cos{k} \end{array} \right]. 
\]
\end{widetext}
The mean fields are determined by self-consistently solving the single-particle
problem of $H_{MF}$ and calculating the corresponding expectation
values according to Eqs. (\ref{chibdel},\ref{chileg}).

Diagonalization of the bosonic part of the
Hamiltonian yields two holon bands. In the ground state of the
system, the holons are assumed to Bose condense into their lowest energy
state.\cite{kotliar} 
Denoting the amplitudes of the lowest holon state on the two legs by $A_a$, the bosonic bond mean fields become
\begin{eqnarray}
B_a &=& 2 \delta \,A^{2}_a, \qquad\quad a=1,2,\label{212} \\\nonumber
B_3 &=& 2 \delta \,A_1\,A_2. 
\end{eqnarray} 
The spinon part of the Hamiltonian can be diagonalized by a 
Bogoliubov transformation, from which the
self-consistent equations for the mean fields $\Delta$ and $\chi$, and the
Lagrange multipliers $\lambda$ are determined numerically.
Furthermore we define the BCS order parameters $\Delta'$,
as\cite{threeleg} 
\begin{eqnarray}\label{BCSmeanfield}
\Delta'_{j,a;j',a'}&:=&\langle
 c_{j,a,\uparrow}c_{j',a'\downarrow}\rangle \\\nonumber 
 &\approx& \langle b^{\dag}_{j,a}b^{\dag}_{j',a'} \rangle\langle
 f_{j,a,\uparrow}f_{j',a'\downarrow}\rangle 
= B_{j,a;j',a'}\Delta_{j,a;j',a'}
\end{eqnarray}
in terms  of the holon and spinon mean fields. For the BCS order
parameters we use the same bond labeling scheme as in
Eq. (\ref{meanfieldlabeling}). 

The redistribution of charge carriers due to the
chemical potential difference is implemented via the holon degrees of
freedom as can be seen in the mean field Hamiltonian. In this way
there is no effect at half-filling. Furthermore, the
constraint and the renormalization of the coupling constants induces a
non-trivial mutual feedback for the charge and spin degrees of freedom.

\subsection{Results}
Within the above mean field description we are able to
 analyze the behavior of the two-leg ladder for different values of the 
doping concentration, $ \delta $, and chemical potential difference,
 $\Delta \mu $. 
In particular, we are interested
in the BCS order parameters as the indication for Cooper paring.
In the following the parameters of the $t$-$J$ model are fixed to isotropic
coupling with 
$t'=t$ and $J'=J=0.5 t$. Furthermore, we restrict
ourselves to the low doping region $\delta<0.25$, where the above spinon-holon
decomposition is expected to be qualitatively reliable.\cite{threeleg}

\begin{figure}[htbp]
\includegraphics[width=\linewidth]{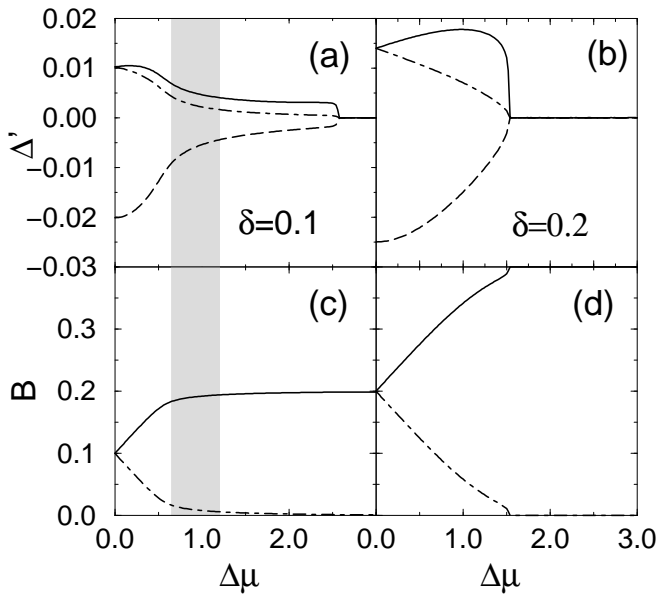}
\caption{BCS mean fields $\Delta'$(a,b) and hole densities $B$ of the $t$-$J$ 
model on a two-leg ladder as functions of the chemical potential difference
$\Delta\mu$, at constant $\delta=0.1$ (a,c), and $0.2$ (b,d), 
for $t'=t$, and $J'=J=0.5t$.
Values for the lower (upper) leg are plotted with solid 
  (dotted-dashed) lines, and the BCS mean field on the rungs using dashed
  lines.} 
\label{figmean1}
\end{figure}
\begin{figure}[htbp]
  \includegraphics[width=\linewidth]{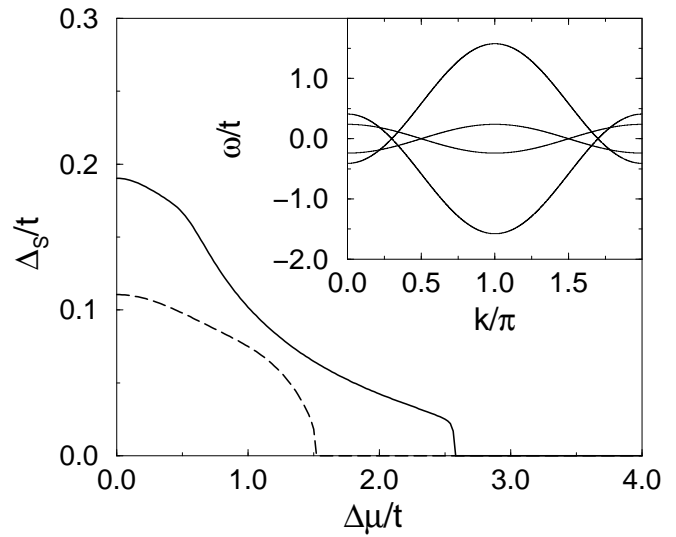}
  \caption{Spinon gap $\Delta_S$ of the $t$-$J$ model on a two-leg ladder as a function of the chemical potential difference
    $\Delta\mu$ at constant hole doping $\delta=0.1$ (solid line) and $\delta=0.2$ (dashed
    line) for $t'=t$, $J'=J=0.5 t$. The inset shows the gapless spinon bands in the normal state at $\delta=0.2$, $\Delta\mu/t=2.0$.}
  \label{figmean3}
\end{figure}
For $\Delta\mu=0$ our calculations agree well with the overall behavior
obtained from similar mean
field calculations using a Gutzwiller-type renormalization
method.\cite{twoleg}  While at half-filling $(\delta=0)$ 
the BCS order parameters vanish, they increase
monotonically with hole doping away from half filling. Their values on
the legs  
coincide ($ \Delta'_1 = \Delta_2' $), whereas a
phase shift of $ \pi $ exists relative to the rung order parameter
$\Delta'_3$, in 
analogy to the d-wave pairing symmetry on the
square lattice version of the doped $ t$-$J$-model.

In order to analyze the influence of a chemical potential
difference between the two legs on the Cooper pairing, we 
follow the
behavior of the BSC mean fields $\Delta'_{1,2,3}$ upon increasing
$\Delta\mu>0$ for two fixed hole concentrations, $\delta=0.1$ and
$\delta=0.2$,
shown in Fig.~\ref{figmean1}~(a,b).
In both cases the chemical potential
difference leads to the reduction and eventual destruction of the 
BCS mean fields. However, 
there is a qualitative difference between the two doping levels.
For $ \delta =0.1 $ a crossover
from a strong to a weak superconducting regime occurs, while no such
regime change takes place for $ \delta=0.2 $.      
The crossover at $ \Delta \mu \approx 0.6 t $ in
Fig. \ref{figmean1}~(a) for $ \delta = 0.1 $ coincides with the almost complete
hole-depletion of the upper leg  in 
Fig. \ref{figmean1}~(c).  
\begin{figure}
\includegraphics[width=\linewidth]{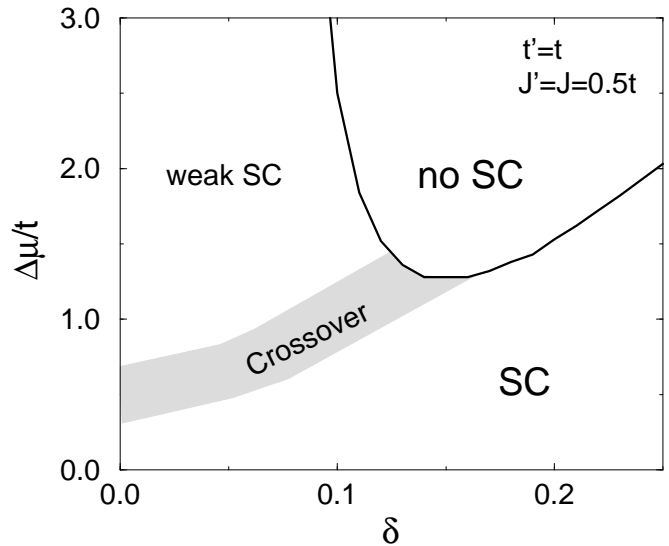}
\caption{Low doping phase diagram of the inhomogeneously doped two-leg
  ladder in the mean 
  field approximation of the  $t$-$J$ model at $t'=t$, $J'=J=0.5 t$. The BCS
  mean fields vanish beyond the solid line. In the low doping region the
  crossover regime connecting the d-wave SC and a regime of reduced BCS mean
  fields is indicated as a shaded area.} 
\label{figmean2}
\end{figure}
This behavior can be understood by the following properties of
doped two-leg ladders. These systems constitute typical examples where
superconductivity originates from a doped RVB phase, which is characterized
within this mean field approach by a finite gap in the spinon spectrum \cite{twoleg}. 
Furthermore, this spinon gap decreases upon doping holes
into a half-filled two-leg ladder \cite{twoleg}. Within the mean field
approximation we can obtain
the spinon gap at finite values of $\Delta\mu$ for the doping
levels considered above.
In Fig.~\ref{figmean3} the
development of the spinon gap upon increasing the chemical potential
difference, $\Delta\mu$, is shown for both $\delta=0.1$, and
$\delta=0.2$. In either case does the inbalance in the distribution of holes
between the two legs lead to an additional reduction of the spinon
gap. This behavior is expected as the RVB phase in the ladder is
dominated by the formation of rung singlet pairs.
Concentrating the holes onto a single leg destroys statistically more rung 
singlets than distributing them equally among both legs. 
However, at $\delta=0.1$, a large
spinon gap is found even for e.g. $\Delta\mu/t=2$, where the upper leg is almost
completely depeted, as seen in Fig.~\ref{figmean1}~(c).
Although the holes are already strongly concentrated onto the lower
leg, the spinon gap is not destroyed until $\Delta\mu$
becomes as large as $\Delta \mu_c \approx 2.6 t$.
Along with the RVB state, d-wave superconductivity thus prevails up to this
critical value of $ \Delta \mu_c $, as seen in  Fig.~\ref{figmean1}~(a).
For the larger doping of $\delta=0.2$ however, 
concentrating the holes onto the lower leg suppresses the spinon gap
completely, therebye destroying the RVB state.
This follows from a comparison of the behavior of the spinon gap in
Fig.~\ref{figmean3} with the corresponding behavior of the charge
distribution shown in Fig.~\ref{figmean1}~(d).
Along with the spinon gap the superconducting state disappears already
at $\Delta\mu_c\approx 1.5 t$.
However, the chemical potential difference which is necessary to
pull the holes onto the lower leg increases upon increasing the hole doping level.

The resulting phase diagram in Fig.~\ref{figmean2} displays 
a peculiar structure. For low-doping concentrations $ \delta <
0.15 $ we observe two regimes, strong and weak superconductivity,
separated by a broad crossover. The crossover region is characterized
by the depletion of holes from the upper leg. 
For larger dopings $ \delta > 0.15 $  only the regime of
strong superconductivity remains, and the RVB state is destroyed 
once the holes are sufficiently unequal distributed among the two legs. 
The mean field solution suggests that there exists a critical
doping $ \delta_c \approx 0.08 $ below which the superconducting state
along with the RVB spin liquid state remains stable for all $ \Delta
\mu $. 

Within the mean field approximation
the non-superconducting phase appears to consist of two independent subsystems.
This can be referred from the inset of Fig.~\ref{figmean3}, which displays the
gapless spinon bands at $\delta=0.2$ and $\Delta\mu/t=2.0$, well inside the normal phase.
The spectrum consists of the spinon bands of a spin-1/2 antiferromagnetic Heisenberg
chain, with nodes at $k=\pm \pi/2$, and two additional bands, 
with nodes at $k_F=\pm\pi/2(1-2\delta)$. These nodes correspond to those of a single
chain Luttinger liquid.
The system in the normal state therefore appear to be separated in 
a $t$-$J$ chain  with  hole doping $2\delta$,
having the properties of a Luttinger liquid, and a
spin-1/2 antiferromagnetic Heisenberg chain.
Furthermore, gapless charge excitations only exist for the
Luttinger liquid. This complete separation is likely an artefact of the mean field
approximation, as in the strong rung coupling limit $ J' \gg J $ the system
is obviously a single chain Luttinger liquid. 
This conclusion results also from the analysis of the
weak-coupling Hubbard model in the next section. 

The mean field description of the $t$-$J$ ladder is in qualitative
agreement with the numerical result of the previous chapter.
In the following section we will analyze the Hubbard model by means of a 
renormalization group treatment in the weak coupling regime, which
reflects the same basic properties of the inhomogeneously doped ladder.

\section{Renormalization group}\label{rg}
In order to complement the analysis of the $t$-$J$ model,
in this section we consider the weakly interacting Hubbard model on a two-leg
ladder. 
A renormalization group (RG) treatment supplemented by Abelian
bosonization allows a detailed analysis of the phase diagram in the weakly
interacting limit,
and a characterization of the various phases in terms of the low-energy
modes. We follow an approach established on  standard two- and N-leg ladder systems (i.e. for $\Delta
\mu=0$).\cite{balents,lin,linso8,ledermann}

In the current case of an inhomogeneously doped two-leg ladder we consider the
weak repulsive limit, $0<U \ll t,t'$ of the Hubbard model,
%\begin{subequations}
\begin{eqnarray}\label{hubbardham}
H & = & - t \sum_{j,a,s}  ( c^{\dag}_{j,a,s} c_{j+1,a,s} +\mathrm{h.c.} )\nonumber \\
& & - t' \sum_{j,s}  ( c^{\dag}_{j,1,s}
c_{j,2,s} + \mathrm{h.c.} )  \\
&& + U \sum_{j,a} c^{\dag}_{j,a,\uparrow} c_{j,a,\uparrow}
c^{\dag}_{j,a,\downarrow}  c_{j,a,\downarrow} \nonumber \\
&& - \sum_{j,a,s} \mu_a c^{\dag}_{j,a,s} c_{j,a,s}, \nonumber 
\end{eqnarray}
%\end{subequations}
with the same notations as is Sec. I. The quadratic part of the Hamiltonian
(\ref{hubbardham}), i.e. Eq. (\ref{hubbardham}) for $U=0$, can be
decoupled via a canonical transformation
\begin{equation}
d^{\dagger}_{j,1/2,s}=\sqrt{\frac{1\mp\Delta\mu/D}{2}}
  c^{\dagger}_{j,2,s}\pm  \sqrt{\frac{1\pm\Delta\mu/D}{2}}  c^{\dagger}_{j,1,s} ,
\end{equation}
where $D=\sqrt{4{t'}^2+\Delta\mu^2}$. These rung operators interpolate smoothly
between the  bonding and anti-bonding combinations at  $\Delta\mu=0$, and the
original fermions for $\Delta\mu/t'\rightarrow \infty$, where
$d^{\dagger}_{j,i,s}\rightarrow c^{\dagger}_{j,i,s}$.
In momentum space two bands 
corresponding to 
$d^{\dagger}_{i,s}(k)$, $i=1,2$
result with dispersions
\begin{equation}\label{hoppingdisperions}
\varepsilon_{1/2}(k)=-2t \cos(k) \pm \frac{1}{2}D-\bar{\mu},
\end{equation}
and a bandwidth $4t$. 

Consider now the effect of the Hubbard interaction term
in Eq. (\ref{hubbardham}).
When both bands are completely separated in energy, only the lower band is filled,
and at half filling ($\delta=0$) a band insulator is obtained which upon doping ($\delta>0$) becomes an
ordinary spin-1/2 Luttinger liquid (LL). 
For $D<4t
\cos^2(\pi\delta)$, both bands are partially filled, and inter-band interaction effects
must be examined. While this proves difficult in general, progress can be made
upon considering the weakly interacting limit.
Since the interest is thus on  the low-energy physics,  the
dispersions (\ref{hoppingdisperions}) can be linearized around the Fermi points $k_{F,i}$, $i=1,2$, determined
by
$\varepsilon_1(k_{F,1})$=$\varepsilon_2(k_{F,2})$ and
$k_{F,1}+k_{F,2}=\pi(1-\delta)$. Furthermore left- and right
movers, $d^{\dagger}_{R/L,i,s}$, are defined with respect to the Fermi level in each band,
$i=1,2$. For
generic (i.e. incommensurate) Fermi momenta the 
interactions consist of intra- and inter-band forward- and Cooper- scattering. These
can be organized in terms of the U(1) and SU(2) current
operators 
\begin{eqnarray}
J_{pij}&=&\sum_s d^{\dagger}_{p,i,s} d_{p,j,s},\nonumber \\
J^{\alpha}_{pij}&=&\frac{1}{2}\sum_{s,s'} d^{\dagger}_{p,i,s}
\sigma^{\alpha}_{s,s'} d_{p,j,s'}  ,
\end{eqnarray}
where $p=R,L$, and the band indices $i,j=1,2$.
The following non-chiral current-current interactions are allowed by symmetry,
\begin{eqnarray}
H_{I}&=&\sum_{i}\int\!\! dx\: (c^{\rho}_{ii} J_{Rii} J_{Lii} - c^{\sigma}_{ii}
{\bf J}_{Rii}\cdot {\bf J}_{Lii}) \nonumber  \\
&+& \sum_{i\neq j} \int\!\! dx\: (c^{\rho}_{ij} J_{Rij} J_{Lij} - c^{\sigma}_{ij}
{\bf J}_{Rij}\cdot {\bf J}_{Lij})\quad  \\
&+& \sum_{i\neq j} \int \!\!dx\: (f^{\rho}_{ij} J_{Rii} J_{Ljj} - f^{\sigma}_{ij}
{\bf J}_{Rii}\cdot {\bf J}_{Ljj}).\nonumber
\end{eqnarray}
In this representation $f$ ($c$) denotes couplings related to forward- (Cooper-)
scattering, and the
symmetry of the inter-band scattering terms under the band exchange is
explicitly taken into account.
Using current algebra and operator product expansions, a one-loop RG flow for the
various couplings can be derived, \cite{balents} which in our notation reads
\begin{eqnarray}\label{RGE}
\frac{d c^{\rho}_{ii}}{dl} &=& -\frac{1}{2v_{\bar{i}}} \left[
  (c^{\rho}_{i\bar{i}})^2 +\frac{3}{16}(c^{\sigma}_{i\bar{i}})^2
  \right],\nonumber \\
\frac{d c^{\sigma}_{ii}}{dl} &=&   -\frac{1}{2v_{\bar{i}}} \left[
  \frac{1}{2}(c^{\sigma}_{i\bar{i}})^2 + 2 c^{\rho}_{i\bar{i}}
  c^{\sigma}_{i\bar{i}}   \right]-\frac{1}{2v_i}(c^{\sigma}_{ii})^2,\nonumber \\
\frac{d c^{\rho}_{12}}{dl} &=& - \sum_{i}\frac{1}{2 v_i} \left[ c^{\rho}_{1i}
  c^{\rho}_{i2} + \frac{3}{16} c^{\sigma}_{1i} c^{\sigma}_{i2}
  \right]\nonumber \\
&&+ \frac{2}{v_1 + v_2}\left[ c^{\rho}_{12} f^{\rho}_{12} + \frac{3}{16}
  c^{\sigma}_{12} f^{\sigma}_{12}\right],\nonumber \\
\frac{d c^{\sigma}_{12}}{dl} &=& - \sum_{i}\frac{1}{2 v_i} \left[ c^{\rho}_{1i}
  c^{\sigma}_{i2} +  c^{\sigma}_{1i}
  c^{\rho}_{i2} + \frac{1}{2} c^{\sigma}_{1i} c^{\sigma}_{i2} \right]
  \\
&&+ \frac{2}{v_1 + v_2}\left[ c^{\rho}_{12} f^{\sigma}_{12} + c^{\sigma}_{12} f^{\rho}_{12} - \frac{1}{2}
  c^{\sigma}_{12} f^{\sigma}_{12}\right], \nonumber \\
\frac{d f^{\rho}_{12}}{dl} &=&\frac{1}{v_1 + v_2}\left[ (c^{\rho}_{12})^2 +
  \frac{3}{16}(c^{\sigma}_{12})^2\right],\nonumber \\
\frac{d f^{\sigma}_{12}}{dl} &=& \frac{1}{v_1 + v_2}\left[ 2 c^{\rho}_{12}
  c^{\sigma}_{12}-\frac{1}{2}(c^{\sigma}_{12})^2-(f^{\sigma}_{12})^2\right],\nonumber 
\end{eqnarray}
where $i=1,2$, $\bar{1}=2$, $\bar{2}=1$, and $v_{i}=2 t \sin(k_{F,i})$ are the
Fermi velocities for the bands. The successive elimination of high frequency
modes is obtained from (\ref{RGE}) by integration along the  logarithmic length scale $l$, related to
an energy scale $E \sim t e^{-\pi l}$ .
The flow equations can be integrated once the
bare values of the couplings are known. For the Hubbard
interaction of Eq. (\ref{hubbardham}) they are obtained as  
\begin{eqnarray}
c^{\sigma}_{11}&=&c^{\sigma}_{22}=4 c^{\rho}_{11}=4 c^{\rho}_{22}=U \left[1+\left(\frac{\Delta\mu}{D}\right)^2\right], \nonumber \\
c^{\sigma}_{12}&=&4 c^{\rho}_{12}=f^{\sigma}_{12}=4
f^{\rho}_{12}=U \left[1-\left(\frac{\Delta\mu}{D}\right)^2\right].
\end{eqnarray}
Increasing $\Delta\mu$ away from zero  can be seen to reduce  the bare inter-band scattering with respect to
intra-band scattering.

Depending on the parameters, integration of the flow equations leads to
different asymptotic behavior, with 
either a flow 
to a finite-valued fix point, or to instabilities characterized by universal ratios
of the renormalized couplings beyond a scale $l^*$, where the most diverging coupling
becomes of  the order the bandwidth.  While the consistency of the one-loop
renormalization group equations is restricted to $l<l^*$, the
asymptotic ratios can be utilized to derive a description of the low-energy physics
of the system within Abelian bosonization.\cite{haldane} Introducing canonically
conjugated bosonic fields $\Phi_{\nu i}$, and $\Pi_{\nu i}$ for the charge and
spin degrees of freedom ($\nu=\rho,\sigma$) on each band $i=1,2$, the
fermionic operators can be represented as 
\begin{equation}
d_{R/L,i,s}=\frac{\eta_{is}}{\sqrt{2\pi\alpha}} \: \mbox{e}^{\: i
    \sqrt{\pi/2}[\pm(\Phi_{\rho i} + s \Phi_{\sigma i}) - (\theta_{\rho i}+
    s\theta_{\sigma i}) ] },
\end{equation}
where $\theta_{\nu i}$ is the dual field of
$\Phi_{\nu i}$, so that $\partial_x \theta_{\nu i}=\Pi_{\nu i}$. The $\eta_{i s}$ are 
Klein factors, ensuring  anticommutation relations, and $\alpha$ is a
short-distance cutoff.\cite{schulz} Using the above representation, the interacting fermionic
Hamiltonian transforms into  a bosonic Hamiltonian,
$H_B=H_q+H_{I}$, containing quadratic terms
\begin{eqnarray}\label{bosonhamq}
H_q&=&  \sum_{\nu,i } \frac{1}{2} \int \!\! dx\left[
  v_i+\frac{c^{\nu}_{ii}}{\beta_{\nu}|\beta_{\nu}| }\right] \partial_x\!
  \Phi_{\nu i}^2 + \left[v_i-\frac{c^{\nu}_{ii}}{\beta_{\nu}|\beta_{\nu}|}
  \right] \Pi_{\nu i}^2 \nonumber \\
 & & + \int \!\! dx\: \frac{f^{\nu}_{12}}{\beta_{\nu}|\beta_{\nu}|} (\partial_x\Phi_{\nu 1}
  \partial_x\Phi_{\nu 2} - \Pi_{\nu 1}\Pi_{\nu 2}),  
\end{eqnarray}
and  sine-Gordon-like interaction terms
\begin{eqnarray}\label{bosonhami}
H_{I}&=&  \int \!\! dx\:  \{ \:c^{\sigma}_{11}\cos{(\sqrt{2} \beta_{\sigma} \Phi_{\sigma_1})} +
c^{\sigma}_{22}\cos{(\sqrt{2} \beta_{\sigma} \Phi_{\sigma_2})} \nonumber \\
& & -4c^{\rho}_{12} \cos{(2 \beta_{\rho}\theta_{\rho -})}
[\cos{(\beta_{\sigma} \Phi_{\sigma -})} - \cos{(\beta_{\sigma}\theta_{\sigma
    -})}] \nonumber \\
& & -c^{\sigma}_{12} \cos{(2\beta_{\rho}\theta_{\rho
    -})}[2 \cos{(\beta_{\sigma}\Phi_{\sigma +})}+
  \cos{(\beta_{\sigma}\Phi_{\sigma -})}] \nonumber \\
& & -c^{\sigma}_{12} \cos{(2\beta_{\rho}\theta_{\rho
    -})}\cos{(\beta \theta_{\sigma -})}  \nonumber \\
& & + 2 f^{\sigma}_{12} \cos{(\beta_{\sigma}\theta_{\sigma -})} \cos{(\beta_{\sigma}
  \Phi_{\sigma +})}\: \} , 
\end{eqnarray}
where $\beta_{\rho}=\sqrt{\pi}$, $\beta_{\sigma}=-\sqrt{4\pi}$, and the fields $\Phi_{\nu\pm}=(\Phi_{\nu 1}\pm \Phi_{\nu
  2})/\sqrt{2}$ and $\Pi_{\nu\pm}=(\Pi_{\nu 1}\pm \Pi_{\nu 2})/\sqrt{2}$ have
been introduced. Upon minimizing the energy in a semiclassical approximation, any coupling that diverges
under the RG flow opens up a gap for a field that is pinned by the corresponding terms in (\ref{bosonhami}).

% Furthermore, the
%  quadratic part of the Hamiltonian, $H_q$, can be diagonalized as 
%\begin{equation}
%\bar{H}_q=  \sum_{\nu,i } \:\frac{u_{\nu i}}{2} \int \!\! dx\: \frac{1}{K_{\nu
%  i}} (\partial_x
%  \bar{\Phi}_{\nu i})^2 +  K_{\nu i} \bar{\Pi}_{\nu i}^2,
%\end{equation}
%where
%$u_{\nu i}$ define the velocities of the low-energy excitations, and the
% Luttinger liquid parameters, $K_{\nu i}$, govern the long-distance
%behaviour of the various correlations functions.\cite{balents} 
%Under the  RG tranformations in the present 
%the diagonalized fields, $\bar{\Phi}_{\mu i}$,  are either 
%rotated  away from the original $\bar{\Phi}_{\mu i}$ with a mixing of the order of $U/t \ll 1$, or
%correspond to the (anti-) symmetric combinations $\Phi_{\mu \pm}$. 

Performing the above procedure, four different
phases  are obtained for the Hamiltonian of Eq. (\ref{hubbardham}), shown in
the ($\delta$, $\Delta\mu)$-plane
for isotropic hopping, $t'=t$, in Fig. 5. 
The 
various phases are labeled according to the number of gapless charge (n) and spin (m) modes by
CnSm.
The different asymptotic regimes of the
RG flow (\ref{RGE}) are related to the phases shown in Fig. 5 as follows:
\begin{figure}
\includegraphics[width=\linewidth]{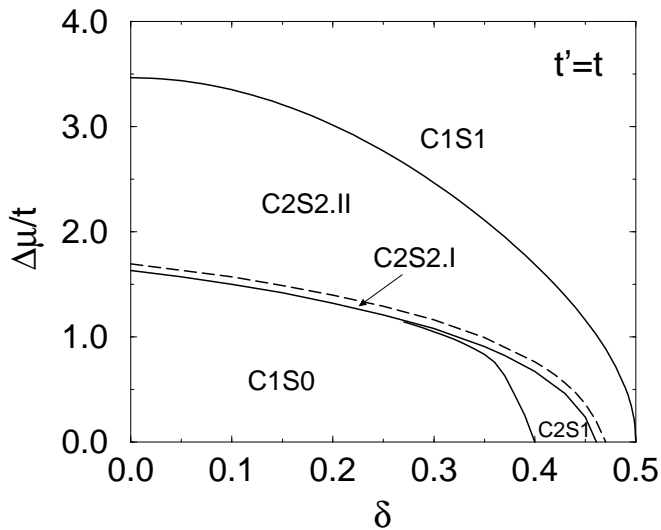}
\caption{Phase diagram of the inhomogeneously doped Hubbard
  model on two-leg ladder in the weakly interacting limit, and for $t'=t$. The
  phases CnSm are labeled according to
  the number of gapless charge (n) and spin (m) modes. Solid lines are
  phase boundaries. The dashed line indicates the crossover from a 
  region with dominant SC correlations in the lower and CDW and SDW
  correlations in the upper band (C2S2.I), to a region dominated by CDW and SDW correlations
  in both bands, for larger values of $\Delta\mu$ (C2S2.II).}
\label{figRG1}
\end{figure}

(C1S1) The single band LL.  This phase with an empty upper
band is labeled C1S1, reflecting the number of gapless modes. For incommensurate
filling the dominant
correlations are charge density waves (CDW) and spin density waves (SDW).\cite{schulz}

(C2S2) The trivial fixed point.
In this regime the couplings stay
of order $U$ under the RG flow, or renormalize to
zero. Therefore no
gap opens in this phase, labeled C2S2.
Furthermore, the two partially filled bands are decoupled in this regime. 
Standard LL theory\cite{schulz}, used to determine the dominant 
 correlations within each band, indicates a crossover between two regions labeled
C2S2.I, and C2S2.II respectively, c.f. Fig. 5. The dominant correlations in the C2S2.II regime are 
 CDW and SDW within both bands,
whereas in the C2S2.I regime the lower band is dominated by superconducting (SC)
fluctuations. 

(C2S1) Single-band superconductivity.
Here, all couplings stay of the order of $U$ or renormalize to zero, except for
$c^{\sigma}_{22} \sim -v_2$. This results in a pinning of the spin
mode of the lower band, and the number of gapless modes reduces to 
C2S1. SC correlations dominate the lower band, and CDW and SDW
correlations the upper band. Furthermore,  inter-band phase coherence is not
established within this regime.
 
(C1S0) Inter-band superconductivity.
In this regime the diverging couplings flow towards the asymptotic ratios
\begin{equation}
4c^{\rho}_{12}=8f^{\rho}_{12}=c^{\sigma}_{12}, \quad c_{11}^{\sigma}/v_1=c_{22}^{\sigma}/v_2.
\end{equation}
From the low-energy effective bosonic Hamiltonian two finite spin gaps are
obtained, and a 
 pinning of the charge mode 
$\theta_{\rho -}=0$, resulting in
phase coherence between the two bands.
The number of gapless modes is reduced to C1S0. 
The remaining total charge mode, ($\Phi_{\rho +}, \theta_{\rho +}$), is
gapless and  exhibits dominant superconducting pairing correlations, with a
sign difference between the bands. This is usually referred to as the d-wave
-like superconducting phase of the two-leg ladder.\cite{balents,lin}

The phase diagram in Fig. 5 confirms the results obtained along the line
$\Delta\mu=0$.\cite{balents,lin} 
But it also indicates a limited stability of the various phases found at
$\Delta\mu=0$ under inhomogeneously doping of the ladder. 
Upon increasing $\Delta\mu$, superconductivity is  gradually suppressed, 
with intermediate phases showing residual superconducting fluctuations.
Consider for example the low doping region  where inter-band d-wave superconductivity
occurs for vanishing $\Delta\mu$. While for small
$\Delta\mu>0$ d-wave superconductivity sustains, 
inter-band phase coherence is lost when $\Delta\mu$ reaches a value of
approximately $1.5 t'$. For larger values of $\Delta\mu$ intra-band
superconductivity persists within the
lower band (which predominately projects onto the lower
leg of the ladder). In the upper band spin fluctuations have become gapless
and SDW and CDW correlations dominate. Further increase of $\Delta\mu$ results
in the suppression of all superconducting
correlations, giving 
rise to two-band LL behavior. In the weakly interacting limit
the upper band is depleted 
for $\Delta\mu/t>3.5$, and a single-band LL,
residing mainly on the lower leg of the ladder, dominates the
large-$\Delta\mu$ regime.  
This progressive reduction of superconducting pairing correlations is also observed
in the finite-temperature phase diagram. In Fig. 6 we show results for $t'=t$,
and $\Delta\mu/t=0.3$ in the $(\delta, T)$-plane. From the renormalization of the energy scale, $E \sim t e^{-\pi l}$, the logarithmic length scale of
Eq. (\ref{RGE}) can be related to a temperature scale $T=E=T_0 e^{-\pi l}$.
While the phases of the system for  $T\rightarrow 0$ are found in accordance with
Fig. 5, the finite temperature phase
diagram  reveals a successive enhancement of superconducting pairing
correlations with decreasing temperature.
Consider again the behavior close to half filling.
At high temperatures the system is dominated by LL behavior in both
bands. Upon decreasing the temperature
gapless superconducting correlations develop within the lower band. At even
lower temperatures, a finite
spin gap opens for the lower band, then finally phase coherent inter-band d-wave
superconductivity emerges, along with the opening of the second spin gap.
Thus in an intermediate temperature regime,
well above the onset of d-wave superconductivity, a single spin gap persists in the
lower band, which is related to the bonding band at small values of $\Delta\mu$. This partial spin gap
formation might be
interpreted as a phenomenon similar to the pseudogap phase in the HTCS
materials.

\begin{figure}[t]
\includegraphics[width=\linewidth]{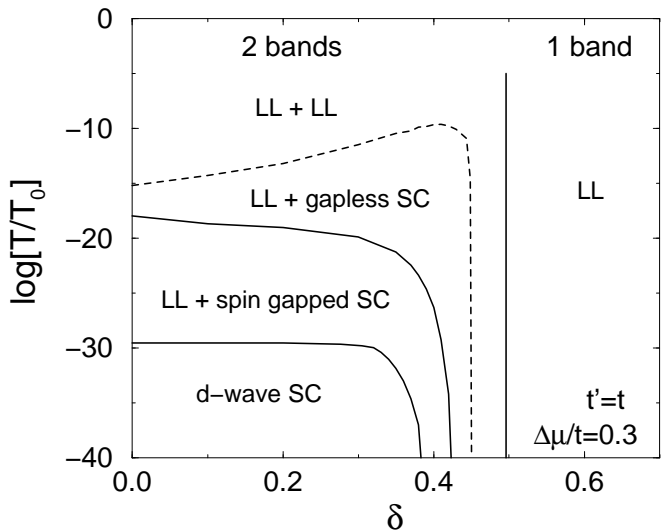}
\caption{Finite Temperature phase diagram of the inhomogeneously doped Hubbard
  model on the two-leg ladder in the weakly interacting limit for $t'=t$, and
  $\Delta\mu/t=0.3$. 
Solid  lines are phase boundaries, whereas the dashed line indicates a
  crossover inside the gapless regime. }
\label{figRG2}
\end{figure}

\section{Discussion and conclusion}
For many years two-leg ladder systems have been prominent model systems for
discussing superconductivity in doped spin liquids.  
In this paper we investigated the stability of the
superconducting phase for inhomogeneous doping by various approaches
which yield an overall similar picture. 

As anticipated from a strong coupling point of view 
a chemical potential difference
between the legs of the ladder acts pair breaking, as could be clearly
demonstrated in numerical exact diagonalization of finite systems. 

The mean field analysis based on the spinon-holon decomposition
suggests that the inbalanced carrier distribution indeed leads to the
suppression of the superconducting state on the doped
ladder. Nevertheless, a more differeniated picture emerges. In the
low-doping region the RVB state remains stable even for large
differences in the chemical potential and supports a weakly
superconducting phase. This RVB phase and the weak superconducting
state do not exist for higher doping concentrations above $ \delta
\approx 0.15 $.  

A modified picture is observed in the renormalization
group treatment of the  weakly interacting Hubbard model on the
two-leg ladder. Also here inhomogenous doping leads to
a suppression of the superconducting phase, a
Luther-Emery-liquid characterized by
one gapless charge mode (C1S0). 
Moreover, an intermediate phase appears which corresponds to a 
single channel being superconducting while a
coexisting channel forms a Luttinger liquid (C2S1).
%Analogous to the mean field result the extent of this phase increases
%towards larger doping 
%concentrations. 
In both the $t$-$J$ and the Hubbard model a phase of
complete destruction of superconducting fluctuations appears for large enough
differences in the 
chemical potential. Within the renormalization group approach
this normal phase is characterized as a single Luttinger liquid state (C1S1).
While this identifies the true low-energy properties of this regime,
the change in 
the spinon spectrum  discussed in Sec. III rather reflects a
short-coming of the 
mean-field solution.

%The $ t$-$J$-model,  despite renormalizations, overestimates the spin
%exchange of the doped system, which is important in stabilizing the
%pairing state.  
%This results in an increase with doping of the stability of the
%superconducting state 
%against the chemical potential difference in the $t$-$J$-model,
%contrasted by a decrease in the Hubbard-model (c.f. Figs. 4 and 6). 

In conclusion we emphasize that inhomogeneous doping of
the two-leg ladder is harmful for the formation of the superconducting
state. Furthermore, it 
can be an interesting tool to access new phases for this type of
electronic systems. 

We acknowledge fruitful discussions with Stephan Haas, Masashige Matsumoto,
and Igor Milat. This work has been financially supported by the Swiss
Nationalfund.


\begin{thebibliography}{99}
\bibitem{footnote} The system with L=9 also exhibits a clear tendency towards
   unbinding, however no level crossing but rather a crossover between the two
   regimes occurs. For $L=11$ a level crossing is found, with the total momentum
   jumping from 0 to a finite value near $0$, whereas in the even systems the
   total momentum changes from 0 to a value near $\pi$. In view of these
   odd-even effect, we restrict the discussion to finite ladders with an even
   number of rungs.
\end{thebibliography}
\end{document}